\documentstyle[12pt]{article}
\begin{document}
\renewcommand{\theequation}{\arabic{section}.\arabic{equation}}
\def\sss{\scriptscriptstyle}
\def\e {{\epsilon}}
\def\ex#1{\langle #1 \rangle }
\def\A {{{\cal A}}}
\def\N {{{\cal N}}}
\def\w {{\omega}}
\def\av#1{\langle  #1 \rangle}
\def\ket#1{|#1 \rangle}
\def\bra#1{\langle #1|}
\def\Ket#1{||#1 \rangle}
\def\Bra#1{\langle #1||}
\def\ov#1#2{\langle #1 | #2  \rangle }
\def\x{\times}
\def\go{\rightarrow  }
\def\rf#1{{(\ref{#1})}}
\def\nn{\nonumber }
\def\be{\begin{equation}}
\def\ee{\end{equation}}
\def\br{\begin{eqnarray}}
\def\er{\end{eqnarray}}
\def\brn{\begin{eqnarray*}}
\def\ern{\end{eqnarray*}}
\def\sixj#1#2#3#4#5#6{\left\{\negthinspace\begin{array}{ccc}
#1&#2&#3\\#4&#5&#6\end{array}\right\}}
\def\isim{\:\raisebox{-0.5ex}{$\stackrel{\textstyle.}{=}$}\:}
\def\lsim{\:\raisebox{-0.5ex}{$\stackrel{\textstyle<}{\sim}$}\:}
\def\gsim{\:\raisebox{-0.5ex}{$\stackrel{\textstyle>}{\sim}$}\:}
\def\bin#1#2{\left(\negthinspace\begin{array}{c}#1\\#2\end{array}\right)}
\def\mat#1#2#3#4{\left(\negthinspace\begin{array}{cc}
#1&#2\\#3&#4\end{array}\right)}
\def\etal{{\it et al., }}
\def\cf{{\it cf }}
\def\ie{{\em i.e. }}
\def\etc{ {\it etc.}}
\def\del {\delta}
\def\d{\dagger}
\def\endauthors{}
\def\authors#1\endauthors{#1}
\begin{titlepage}
\pagestyle{empty}
\baselineskip=21pt
\begin{center}
{\large{\bf Self Consistent Random Phase Approximation
within the O(5) model and Fermi transitions}}
\end{center}
\vskip .1in

\authors
\centerline{ F. Krmpoti\'{c}$^{a\dagger}$, E.J.V. de Passos$^{b\dagger\dagger}$,
D.S. Delion$^{c}$,
J. Dukelsky$^{d}$ and P. Schuck$^{e}$}
\vskip .05in
\centerline{\it ${}^a$ Departamento de F\'\i sica, Facultad de Ciencias
Exactas}
\centerline{\it Universidad Nacional de La Plata, C. C. 67, 1900 La Plata,
Argentina.}
\vskip .05in
\centerline{\it ${}^b$  Instituto de F\'\i sica, Universidade de
S\~{a}o Paulo,}
\centerline{\it C.P. 66318, 05389-970 S\~{a}o Paulo, Brasil.}
\vskip .05in
\centerline{\it $^{c}$ Institute of Physics and Nuclear Engineering}
\centerline{\it POB MG6 Bucharest, Romania}
\vskip .05in
\centerline{\it $^{d}$ Instituto de Estructura de la Materia}
\centerline{ \it Serrano 123, 28006 Madrid, Spain}
\vskip .05in
\centerline{\it $^{e}$ Institut des Sciences Nucleaires}
\centerline{\it 53 Avenue des Martyrs, F-38026 Grenoble Cedex, France}
\endauthors


\centerline{ {\bf Abstract} }
Self Consistent Quasiparticle Random Phase Approximation (SCQRPA)
is considered in application to the Fermi transitions within the O(5)
model. It is demonstrated that SCQRPA improves on renormalized QRPA (RQRPA),
a method that has recently become rather popular in this context.
The analytical form of the SCQRPA vacuum is used to evaluate all
the matrix elements. The SCQRPA results
show a general trend similar to the exact solutions.
The necessity to change the single particle basis beyond
the transition point, and to include the
proton-proton and neutron-neutron channels in the QRPA operator,
in addition to the proton-neutron
one, is pointed out.

\baselineskip=18pt
\bigskip
\noindent
\\ $^{\dagger}$Fellow of the CONICET from Argentina.
\\ ${}^{\dagger\dagger}$Fellow of the CNPq from Brazil.
\end{titlepage}
\newpage
\section {Introduction}
\label{sec1}

Since Vogel and Zirnbauer \cite{Vog86} and Cha \cite{Cha87} have
discovered the importance
of the particle-particle force in the $S=1$, $T=0$ channel,
the quasiparticle random phase approximation (QRPA) has been the most
frequently used nuclear structure method for evaluating double beta
($\beta\beta$)
rates both for the two-neutrino decay mode ($\beta\beta_{2\nu}$) and
for the neutrinoless mode ($\beta\beta_{0\nu}$). The general feature
of this method is that the resulting nuclear matrix elements ${\cal
M}_{2\nu}$ and ${\cal M}_{0\nu}$ turn out to be highly sensitive to
the particle-particle force
\cite{Vog86,Civ87,Eng88,Mut88,Hir90,Krm93,Krm94}.
Moreover, very close to its
physical value the QRPA collapses.  One thus may suspect that this method
yields relatively small values of ${\cal M}_{2\nu}$ simply because the
approximation breaks down. In other words, the smallness of ${\cal
M}_{2\nu}$ in the QRPA could be just an artifact of the model.

In the past few years, several modifications of the QRPA have been
proposed to improve on the above mentioned behaviour
\cite{Rad91,Rad93,Kuo92,Krm93a,Che93}. Yet, the one that has received
major attention lately is the so-called renormalized QRPA (RQRPA)
\cite{Toi95,Krm96,Sch96,Krm97,Toi97,Sim97,Mut97,Eng97}.
The new ingredient that is brought up is the effect of the ground state
correlations (GSC) in the QRPA equations of motion (EOM) i.e.
while in the standard QRPA the correlated ground state is approximated
by the BCS vacuum, in the RQRPA the GSC are partially taken into account.
Let us recall that the GSC are fully incorporated only in the
self-consistent RPA (SCRPA) theory \cite{Duk90,Duk98}, which simultaneously
leads to a coupling of the single-particle field to the RPA excitations.
The same formalism was also generalized to the superconducting systems
\cite{Duk96}, and when applied in this context will be referred to as the
 self-consistent QRPA (SCQRPA).

The RQRPA generally yields better results than the QRPA, in the sense that the
instability is avoided, and the matrix elements ${\cal M}_{2\nu}$ and
${\cal M}_{0\nu}$ are somewhat less sensitive to the particle-particle
interaction. Yet, the charge-exchange sum rules \cite{Boh75}, both for Fermi
(F) and Gamow-Teller (GT) transitions,  are  significantly violated within
the RQRPA \cite{Krm96,Krm97,Toi97}.
It has been shown \cite{Krm97} that the origin of this violation is the
omission of the contribution of the so-called scattering (anomalous) terms
to these sum rules.
A possible way to solve this problem has been suggested in ref. \cite{Duk96},
where the inclusion of these terms in the definition of the
excitation operators is proposed.

To explore the effect of the GSC on the $\beta\beta$ matrix elements,
several applications of the RQRPA have recently been performed in the framework
of the exactly solvable O(5) model \cite{Hir96,Sam97}, which allows for a
direct comparison between the exact and approximate results.
A similar exercise has also been done within the SCQRPA \cite{Del97},
but by solving the corresponding EOM in an approximate way.
For that reason, and in contrast with experience with the SCQRPA approach
\cite{Duk90,Duk98,Kru94}, the excitations energies and matrix elements for
the $\beta \beta$ decay were not well reproduced.

In this paper we resume the study performed in ref. \cite{Del97}, but
avoiding the approximations done there.
With the aim to clarify the physical meaning of the model parameters,
a brief introduction of the O(5) model is presented in Section \ref{sec2}.
The full SCQRPA equations for the O(5) model, with the simplifying
condition of decoupling the charge-exchage mode from the conserving ones,
are derived in Section \ref{sec3}.
These equations are solved self-consistently with the generalized BCS
equations, which implies  the coupling of the quasi-particle mean-field to
the charge-exchage excitation.
The task has been facilitated by the circumstance that it was possible to
construct explicitly the correlated ground state wave function for the
schematic O(5) Hamiltonian.
In Section \ref{sec4} the RRPA is formulated as the limit of the SCQRPA when
the ground state two-body density is approximated by products of one-body
densities.
Numerical results for the Fermi transitions and the corresponding
$\beta\beta$ decay matrix elements, obtained within different approximations,
are compared with each other and confronted with exact results in Section
\ref{sec5}. Summarizing conclusions are given in Section \ref{sec6}. Finally,
some details of the calculations are displayed in the appendices.

\section {The Model}
\label{sec2}

In the model, protons and neutrons occupy only one level and the Hamiltonian
\br
H &=&\e_pN_{pp}+\e_nN_{nn}
-\frac{G_p}{4}P_{pp}^{\d}P_{pp}-\frac{G_n}{4} P_{nn}^{\d}P_{nn}
\nonumber \\
&+&\frac{1}{4}\left[F_{pn}\left(N_{pn}N_{np}+N_{np}N_{pn}\right)
-G_{pn}\left(P_{pn}^{\d}P_{pn}+P_{np}^{\d}P_{np}\right)\right],
\label{2.1} \er
is such that it can be written in terms of generators of O(5) \cite{Del97}
\br
N_{ab} &=&\sum_mc_{a m}^{\d}c_{b m},
\nn\\
P_{ab}^{\d}&=& P_{ba}^{\d}=\sum_m (-1)^{j+m}c_{a m}^{\d}{c}_{b-m}^{\d},
\nn\\
P_{ab}&=&\left(P_{ab}^{\d}\right)^{\d},
\label{2.2}\er
where $ab$ stands for $pp$, $nn$, $pn$ and $np$.

To see the physical meaning of the model parameters, we rewrite
the model Hamiltonian \rf{2.1} in the coupled basis
\be
C^{\d}_{ab;JM}=\left(c^{\d}_{a}\otimes c^{\d}_{b}\right)_{JM},
\label{2.3}\ee
\ie
\br
H &=& \sum_{a=p,n}\left(\e_p+\frac{F_{pn}}{4}\right)N_{aa}
-\frac{\Omega}{2}\sum_{a=p,n}G_aC^{\d}_{aa;00}C_{aa;00}
\nonumber \\
&-&\Omega G_{pn}C^{\d}_{pn;00}C_{pn;00}
+\frac{F_{pn}}{2} \sum_{JM}(-1)^JC^{\d}_{pn;JM}C_{pn;JM}
\label{2.4} \er
and compare it with the standard one-level shell model Hamiltonian:
\br
H^{\rm {\sss SM}} &=&\sum_{a=p,n}\e^{\rm {\sss SM}}_a N_{aa}
+\frac{1}{4}\sum_{JM;a=p,n}[1+(-1)^J]
\frac{\Bra{j^2_a; J}v\Ket{j^2_a;J}}{\sqrt{2J+1}}
C^{\d}_{aa;JM} C_{aa;JM}
\nonumber \\
&+&\sum_{JM}\frac{\Bra{j_pj_n; J}v\Ket{j_pj_n;J}
+(-1)^J\Bra{j_pj_n; J}v\Ket{j_nj_p;J}}
{\sqrt{2J+1}} C^{\d}_{pn;JM} C_{pn;JM}.
\label{2.5} \er
We conclude that the $\e_a^{\rm {\sss SM}}$ are related with $\e_a$ as
\be
\e_a^{\rm {\sss {\sss SM}}}=\e_a+\frac{F_{pn}}{4},
\label{2.6}\ee
and that the particle-particle matrix elements for the Hamiltonian \rf{2.1} are:
\be
G(aaaa;J)\equiv[1+(-1)^J]\frac{\Bra{j^2_a; J}v\Ket{j^2_a;J}}{\sqrt{2J+1}}
=-2\Omega G_a\del_{J0},
\label{2.7}\ee
and
\br
G(pnpn;J)&\equiv&\frac{\Bra{j_pj_n; J}v\Ket{j_pj_n;J}
+(-1)^J\Bra{j_pj_n; J}v\Ket{j_nj_p;J}}{\sqrt{2J+1}}
\nn\\
&=&-\left(\Omega G_{pn}\del_{J0}-(-1)^J\frac{F_{pn}}{2}\right),
\label{2.8}\er
with $2\Omega=2j+1$.
Therefore like particles interact only in the $J=0$ channel (pairing term).
But, the unlike particles, besides having a pairing term, also interact
in $J\ne 0$ channels. In particular, for odd $J$, the Hamiltonian \rf{2.1}
simulates the $T=0$ pairing. Moreover, it is charge dependent, and the limit
of charge independence is achieved when $\e_p=\e_n$, $G_p=G_n=G_{pn}$
and $F_{pn}=0$.

For further use it is convenient to define the particle-hole matrix elements
\footnote{The particle-hole matrix elements are defined as \cite{Raj69}:
\[
F(adcb;J)=-\sum_{J'}(2J'+1)\sixj{a}{b}{J'}{c}{d}{J}G(abcd;J').
\]}
\be
F(aaaa;J)=-(-1)^{J}G_a
\label{2.9}\ee
\be
F(pnpn;J)=\left(\Omega F_{pn}\del_{J0}-(-1)^J\frac{G_{pn}}{2}\right),
\label{2.10}\ee
\be
F(ppnn;J)=-\frac{1}{2}\left(F_{pn}+(-1)^JG_{pn}\right).
\label{2.11}\ee
It might be worth noting that to make the Hamiltonian \rf{2.1} charge
conserving, even when $F_{pn}\ne 0$, one has to do the replacement:
\be
H\go H+\frac{F_p}{4}N_{pp}N_{pp}+\frac{F_n}{4} N_{nn}N_{nn}.
\label{2.12}\ee
Now
\br
G(aaaa;J)&=&-2G_a\Omega\delta_{J0}+\frac{F_a}{2}(1+(-1)^J)
\nn\\
F(aaaa;J) &=&-(-1)^JG_a+\frac{F_a}{2}\left(2\Omega\del_{J0}-1\right)
\label{2.13}\er
and the charge independence is fulfilled for $\e_p=\e_n$, $G_p=G_n=G_{pn}$
and $F_p=F_n=F_{pn}$. In this case the following relations are valid:
\br
G(aaaa;0)&=&2G(pnpn;0),
\nn\\
F(aaaa;0)&=&F(pnpn;0)+F(ppnn;0).
\label{2.14}\er

\setcounter{equation}{0}
\renewcommand{\theequation}{\arabic{section}.\arabic{subsection}.\arabic{equation}}

\section {The SCQRPA Equations}
\label{sec3}

\subsection{General Formulation}
\label{sec3.1}

The "nuclei" to be considered here are usually superfluid both for protons
and for neutrons. We therefore make first a Bogoljubov transformation to
quasiparticle operators $a^{\d}_{a m}$,
\be
c_{a m}^{\d}=u_a a_{a m}^{\d}+v_a(-1)^{j+m}a_{a,-m},~~~~a=p,n.
\label{3.1.1}\ee
The SCQRPA states are constructed by the action of the excitation operators
\footnote{
In recent works \cite{Duk98,Duk96} it was proposed to also include anomalous
terms $a^{\dagger}a$ in the excitation operator. Though this potentially is of
great importance we do not consider this extension here (see also the
discussion at the end of this paper). }
\be
\Gamma^{\d}_\nu=\sum_{\tau =pp,nn,pn}
X_\tau^\nu{\tilde \A}_\tau^{\d}
-Y_\tau^\nu{\tilde \A}_\tau,~~~~~~~\nu=1,2,3
\label{3.1.2}\ee
on the SCQRPA ground state $\ket{0}$,
\be
\ket{\nu}=\Gamma_\nu^{\d}\ket{0},
\label{3.1.3}\ee
with the vacuum condition
\be
\Gamma_\nu\ket{0}=0.
\label{3.1.4}\ee
The operators
\be
{\tilde \A}_\tau^{\d}=\frac{\A_{ab}^{\d}}{\sqrt{2\Omega(1+\del_{ab})}},
\label{3.1.5}\ee
which appear in eq. \rf{3.1.2} create  $pp$, $nn$ and $pn$
quasiparticle pairs with $J=0$:
\be
\A_{ab}^{\d}= \sum_m (-1)^{j+m}a_{a m}^{\d}a_{b,-m}^{\d},~~~~~a,b=p,n.
\label{3.1.6}\ee
and therefore the ansatz \rf{3.1.2} can be considered as a Bogoljubov
transformation between fermion pair operators.

The normalization condition
\be
\ov{\nu}{\nu'}=\bra{0}\left[\Gamma_{\nu'},\Gamma_\nu^{\d}\right]\ket{0}
=\del_{\nu\nu'},
\label{3.1.7}\ee
gives
\be
\sum_{\tau'\tau}\left(X^{\nu'*}_{\tau'}D_{\tau'\tau}X_\tau^\nu
-Y^{\nu'*}_{\tau'}D_{\tau'\tau}^*Y_\tau^\nu\right) =\del_{\nu\nu'},
\label{3.1.8}\ee
where
\be
D_{\tau'\tau}=\bra{0}\left[{\tilde \A}_{\tau'},{\tilde \A}_\tau^{\d}\right]
\ket{0}.
\label{3.1.9}\ee
The SCQRPA equations have then the form:
\be
\mat{A}{B}{B^*}{A^*}\bin{X}{Y}= \w\mat{D}{0}{0}{-D^*}\bin{X}{Y},
\label{3.1.10}\ee
where
\br
A_{\tau\tau'}&=&\bra{0}\left[{\tilde \A}_{\tau},{\cal H},
{\tilde \A}_{\tau'}^{\d}\right]\ket{0} =A^*_{\tau'\tau},
\nn\\
B_{\tau\tau'}&=&-\bra{0}\left[{\tilde \A}_{\tau},{\cal H},
{\tilde \A}_{\tau'}\right]\ket{0} =B_{\tau'\tau},
\label{3.1.11}\er
are the symmetrised double commutators defined by Rowe \cite{Row68} with
\be
{\cal H}=H-\lambda_pN_{pp}-\lambda_nN_{nn}.
\label{3.1.12}\ee

The SCRPA equations are still incomplete because the mean field
parameters $ u_{a}$, $v_{a}$, and $\lambda_a$ are so far
undetermined. This general problem of SCRPA has been solved very
satisfactorily in \cite{Duk96} where it has been shown that the mean field
parameters (the single particle basis in a non-superfluid system
and/or the amplitudes $ u_{a}$, $v_{a}$ in a superfluid one)
can be determined in minimizing the SC(Q)RPA ground state energy with
respect to these parameters. This minimization turns out to be
equivalent to \cite{Duk96}:
\be
\bra{0}[{\cal H},\Gamma^{\d}_\nu]\ket{0}=0;~~~~~\nu=1,2,3,
\label{3.1.13}\ee
which with \rf{3.1.4} is a very natural equation from the EOM point of view.

The eq. \rf{3.1.13} is equivalent to
\be
\bra{0}[{\cal H},{\tilde \A}_\tau^{\d}]\ket{0}=0,
\label{3.1.14}\ee
which, upon replacing $\ket{0}$ by the mean field ground state, turns out to
be identical to the usual BCS equation.
\footnote
{These generalized mean field equations also serve to make the double
commutators \rf{3.1.11} symmetric. In the O(5) model the symmetry of the double
commutators is verified (see eqs. \rf{3.5.9}).}

The chemical potentials $\lambda_{n,p}$
are determined from the usual particle number condition, \ie  it is imposed
that the SCQRPA ground state $\ket{0}$ has, on average, the correct number of
protons ($Z={\rm N}_p$) and neutrons ($N={\rm N}_n$)
\be
\bra{0}N_{aa}\ket{0}={\rm N}_a.
\label{3.1.15}\ee
The set of equations \rf{3.1.4}, \rf{3.1.10}, \rf{3.1.11} and \rf{3.1.15}
have to be solved simultaneously, together with the conditions \rf{3.1.14}.
This generally makes the problem to be very demanding
computationally. Indeed the problem has only been tackled so far
within very simple schematic models \cite{Duk96}.
Here we are going to work out the SCQRPA equations of the O(5) model, under
some simplifying conditions. That is, we assume that the charge-exchange mode,
\be
\Gamma^{\d}_{pn}=X_{pn}\frac{\A^{\d}_{pn}}{\sqrt{2\Omega}}
-Y_{pn}\frac{\A_{pn}}{\sqrt{2\Omega}},
\label{3.1.16}\ee
decouples from the modes involving like particles pairs,
\be
\Gamma^{\d}_\nu=\sum_{a=p,n}X^\nu_{a a}\frac{\A^{\d}_{a a}}{2\sqrt{\Omega}}
-Y^\nu_{a a}\frac{\A_{a a}}{2\sqrt{\Omega}}.
\label{3.1.17}\ee
Since both the Hamiltonian and the Bogoljubov transformation are charge
conserving, the decoupling happens automatically for a SCQRPA vacuum that is a
superposition of states with even number of protons and neutrons. This is
achieved by taking into account only the correlations due to the $pn$ pairs.
In other words, the SCQRPA vacuum is determined only by the equation
\be
\Gamma_{pn}\ket{0}=0.
\label{3.1.18}\ee
Within this framework, and since
\be
\bra{0}[{\cal H},\A_{pn}^{\d}]\ket{0}=0
\label{3.1.19}\ee
is trivially satisfied, the eqs. \rf{3.1.15} for the mean field reduce to
\be
\bra{0}[{\cal H},\A_{pp}^{\d}]\ket{0}=
\bra{0}[{\cal H},\A_{nn}^{\d}]\ket{0}=0,
\label{3.1.20}\ee
and the number condition \rf{3.1.14} has to be obeyed.

In summary, since in this paper we are interested in investigating the
Fermi transitions in this schematic model, our equations are: \rf{3.1.16}
for the excitation operator, \rf{3.1.18} for the SCQRPA vacuum, and
\rf{3.1.14} and \rf{3.1.19} for the mean field.

\setcounter{equation}{0}

\subsection{Hamiltonian in the quasiparticle basis}
\label{sec3.2}

The first step in solving the above  equations is to rewrite the
Hamiltonian \rf{3.1.12} in the quasiparticle basis. Using the Bogoljubov
transformation, ${\cal H}$ can be put in normal order with respect to the
quasiparticle vacuum. Five type of terms occur,
\be
{\cal H} ={\cal H}^{00}+{\cal H}^{11}+{\cal H}^{20}+
{\cal H}^{22}+{\cal H}^{31}+{\cal H}^{40},
\label{3.2.1}\ee
and they can be expressed in terms of the generators of an O(5) algebra
$\A_{ab}^{\d}$,
$\A_{ab}=\left(\A_{ab}^{\d}\right)^{\d}$ and
\be
\N_{ab} =\sum_ma_{a m}^{\d}a_{b m},
\label{3.2.2}\ee
analogous to those given by \rf{2.2}, but now in the quasiparticle basis.
The first term
\begin{eqnarray}
{\cal H}^{00} &=&\sum_{a=p,n}\left\{\Omega\left[
2\left(\e_a^{(0)}-\lambda _a\right)v_a^2-G_a v_a^4\right]
-G_a\Omega^2u_a^2v_a^2\right\}
\nn\\
&-&\Omega(G_{pn}+F_{pn})v_p^2v_n^2,
\label{3.2.3}\er
is the BCS ground state energy, and the second one
\be
{\cal H}^{11}=E_p\N_{pp}+E_n\N_{nn},
\label{3.2.4}\ee
is the one-quasiparticle Hamiltonian, where
\be
E_a= (\epsilon_{a}^{(s)}-\lambda_{a})(u_{a}^{2}-
v_{a}^{2})+2\Omega v_{a}^{2}u_{a}^{2}G_{a},
\label{3.2.5}\ee
are the quasiparticle energies, in the BCS limit. Here
\be
\epsilon_{a}^{(s)}
=\epsilon_{a}^{\sss SM}+\mu_a,
\label{3.2.6}\ee
are the single particle energies, corrected by
the self-energy terms \cite{Krm97}
\br
\mu_p&=&-G_pv^2_p-\frac{1}{2}(G_{pn}+F_{pn})v_n^2,
\nn\\
\mu_n&=&-G_nv^2_n-\frac{1}{2}(G_{pn}+F_{pn})v_p^2.
\label{3.2.7}\er
The remaining quantities are
\br
{\cal H}^{20}&=&\frac{1}{2}\sum_{a}\left[2(\epsilon_{a}^{(s)}
-\lambda_{a})u_{a}v_{a}-\Omega
G_{a}u_{a}v_{a}(u_{a}^{2}-v_{a}^{2})\right]
({\cal A}_{aa}^{\dagger}+{\cal A}_{aa}),
\label{3.2.8}\er
\br
{\cal H}^{40}&=&\frac{1}{4}\sum_{a}G_a u_{a}^2v_{a}^2
({\cal A}_{aa}^{\dagger} {\cal A}_{aa}^{\dagger}
+{\cal A}_{aa}{\cal A}_{aa})
\nn\\
&+&\frac{1}{2}(G_{pn}+F_{pn})u_nv_nu_pv_p
({\cal A}_{pn}^{\dagger} {\cal A}_{pn}^{\dagger}
+{\cal A}_{pn}{\cal A}_{pn}),
\label{3.2.9}\er
\br
{\cal H}^{31}&=&\frac{1}{2}\sum_{a}G_a u_{a}v_{a}(u_a^2-v_a^2)
({\cal A}_{aa}^{\dagger} {\cal N}_{aa}
+{\cal N}_{aa}{\cal A}_{aa})
\nn\\
&+&\frac{1}{2}(G_{pn}+F_{pn})\left[u_nv_n(u_p^2-v_p^2)
({\cal A}_{pn}^{\dagger}{\cal N}_{np} +{\cal N}_{pn}{\cal A}_{pn})
\right.
\nn\\
&+&\left.u_pv_p(u_n^2-v_n^2) ({\cal A}_{pn}^{\dagger}{\cal N}_{pn}
+{\cal N}_{np}{\cal A}_{pn}) \right],
\label{3.2.10}\er
and
\br
{\cal H}^{22}&=&-\frac{1}{4}\sum_{a}G_a (u_a^4+v_a^4)
{\cal A}_{aa}^{\dagger} {\cal A}_{aa}
-\sum_{a}G_a u_a^2v_a^2 ({\cal N}_{aa}^2- {\cal N}_{aa})
\nn\\
&+&\frac{1}{2}[F_{pn}(u_p^2v_n^2+u_n^2v_p^2)
-G_{pn}(u_p^2u_n^2+v_p^2v_n^2)] {\cal A}_{pn}^{\dagger}{\cal A}_{pn}
\nn\\
&-&\frac{1}{2}(G_{pn}+F_{pn})u_pv_pu_nv_n ({\cal N}_{pn}^2 +{\cal N}_{np}^2)
\nn\\
&+&\frac{1}{4}[F_{pn}(u_p^2u_n^2+v_p^2v_n^2)
-G_{pn}(u_p^2v_n^2+u_n^2v_p^2) ]
\nn\\
&\x&({\cal N}_{pn}{\cal N}_{np}+{\cal N}_{np}{\cal N}_{pn}
-{\cal N}_{nn}-{\cal N}_{pp}).
\label{3.2.11}\er

\setcounter{equation}{0}

\subsection{The SCQRPA vacuum}
\label{sec3.3}

The correlated vacuum is defined by eq. \rf{3.1.18}, which is equivalent
to
\be
(\A_{pn}- z\A_{pn}^{\d})\ket{0}=0;~~~z=\frac{Y}{X}.
\label{3.3.1}\ee
From the vacuum condition it is very difficult to find an explicit
expression for the SCQRPA ground state $\ket{0}$ in the general case.
Yet, because of the simplicity of the model used here, the eq. \rf{3.3.1}
can be solved exactly, and one obtains

\be
\ket{0}=\sqrt{N_0}\sum_{l=0}^{\Omega}
a_lz^l\left(\A^{\d}_{pn}\right)^{2l}\ket{BCS}=
\sqrt{N_0}\sum_{l=0}^{\Omega}\alpha_lz^l\ket{2l},
\label{3.3.2}\ee
with $\ket{BCS}$ being the quasiparticle vacuum and
\be
N_0=\left( \sum_{l=0}^{\Omega}\alpha_l^2z^{2l}\ov{2l}{2l}\right)^{-1},
\label{3.3.3}\ee
\be
\alpha_l=(2\Omega)^l\frac{\Omega!}{(2\Omega)!}\frac{(2\Omega-2l))!}
{l!(\Omega-l)!} ;~~~~~~~~~\ov{k}{k}=\frac{k!(2\Omega)!}{(2\Omega-k)!
(2\Omega)^k}.
\label{3.3.4}\ee
Thus, the SCQRPA vacuum is a superposition of states with equal even number
of neutron and proton quasiparticles, and as such a superposition of states
with even number of protons and neutrons. To solve the SCQRPA equations we
have to evaluate the expectation values of the generators $\A_{ab}^{\d}$,
$\A_{ab}$ and $\N_{ab}$ and
their bilinear combinations as well. The results are given in the Appendix A.

\setcounter{equation}{0}

\subsection{Generalized BCS equations}
\label{sec3.4}

The only undetermined quantities are the $u_a$'s and $v_a$'s of the BCS state
and the chemical potentials. They are fixed by the generalized BCS equations
\rf{3.1.19} and the number conditions \rf{3.1.14}. The first ones yield
\br
\bra{0}[{\cal H}^{20}+{\cal H}^{31},\A_{pp}^{\d}]\ket{0}&=&0,
\nn\\
\bra{0}[{\cal H}^{20}+{\cal H}^{31},\A_{nn}^{\d}]\ket{0}&=&0,
\label{3.4.1}\er
and from the second ones we obtain
\footnote{
This generalized particle number equation has for the first time been
incorporated into an extended RPA calculation in refs. \cite{Krm96,Duk96}
though this had not explicitly been mentioned in the latter
of the two references.}
\be
{\rm N}_a=2\Omega v_a^2+(u_a^2-v_a^2)\bra{0}\N_{aa}\ket{0},~~~a=p,n.
\label{3.4.2}\ee
When the SCQRPA vacuum $\ket{0}$ is replaced by the quasiparticle vacuum
$\ket{BCS}$, the eqs. \rf{3.4.1} and \rf{3.4.2} reduce to the usual BCS
equations
\br
\bra{BCS}[{\cal H}^{20},\A_{pp}^{\d}]\ket{BCS}&=&
\left[2(\e_p^{(s)}-\lambda_p)-G_p\Omega(u_p^2-v_p^2)\right]u_pv_p=0,
\nn\\
\bra{BCS}[{\cal H}^{20},\A_{nn}^{\d}]\ket{BCS}&=&
\left[2(\e_n^{(s)}-\lambda_n)-G_n\Omega(u_n^2-v_n^2)\right]u_nv_n=0,
\label{3.4.3}\er
with
\be
v_{a}^{2}=\frac{{\rm N}_{a}}{2\Omega};~~~~~~~~~~~~~~~ a=p,n.
\label{3.4.4}\ee
The SCQRPA correlations lead to the following generalized BCS equations:
\be
2\xi_au_av_a+\Delta_a(v_a^2-u_a^2 ) =0;~~~~~u^2_a+v^2_a=1,
\label{3.4.5}\ee
with the standard solution for $u^{\prime }s$ and $v^{\prime }s$
\br
v^2_a={1\over 2}\left(1-{\xi_a\over \sqrt{\xi_a^2+\Delta_a^2}}\right),
\nn\\
u^2_a={1\over 2}\left(1+{\xi_a\over \sqrt{\xi_a^2+\Delta_a^2}}\right),
\label{3.4.6}\er
where
\br
\xi _p&=&\frac{1}{\Omega}(\epsilon^{(s)}_p-\lambda_p)
\bra{0}(\Omega-{\cal N}_{pp})\ket{0}
\nn\\
&-&\frac{1}{4\Omega}\left( F_{pn}+G_{pn}\right) \left( v_n^2-u_n^2\right)
\bra{0}\left({\cal A}_{pn}^{\dagger }{\cal A}_{pn}-
{\cal N}_{np}{\cal N}_{pn}\right)\ket{0},
\label{3.4.7}\er
and
\br
\Delta_p&=&\frac{1}{2\Omega}G_p u_pv_p\bra{0}\left[2(\Omega
-{\cal N}_{pp})^2-{\cal A}_{pp }^{\dagger }{\cal A}_{pp}\right]\ket{0}
\nn\\
&-&\frac{1}{2\Omega}\left( F_{pn}+G_{pn}\right)
u_nv_n\bra{0}{\cal A}_{pn}^{\dagger }{\cal A}_{pn}^{\dagger }\ket{0},
\label{3.4.8}\er
are, respectively, the renormalized single particle energy and gap parameter
for protons (and similarly for neutrons).
One remarks that formally the generalized BCS equations have the
same structure as the ordinary ones. However proton and neutron
equations are coupled, the pairing interaction is renormalized by the
two body densities (screening) and also the single particle
energies are coupled back to the correlations.

From the number of particle condition \rf{3.4.2} it follows immediately
that the BCS amplitudes \rf{3.4.6} can be expressed as:
\br
v_a^2&=&\frac{1}{2}\left(1-\frac{\Omega-{\rm N}_a}
{\Omega-\bra{0}\N_{aa}\ket{0}}\right),
\nn\\
u_a^2&=&\frac{1}{2}\left(1+\frac{\Omega-{\rm N}_a}
{\Omega-\bra{0}\N_{aa}\ket{0}}\right)=1-v_a^2.
\label{3.4.9}\er
It should be stressed that these expressions, which generalize
\rf{3.4.4} are independent of \rf{3.4.6} and
are equivalent to the eq. \rf{3.4.2}, which fixes the number of particles.
They allow us to express directly the quantities entering
the BCS eqs. \rf{3.4.5} in terms of the one-body densities and
to solve them.

\setcounter{equation}{0}

\subsection{The SCQRPA equations}
\label{sec3.5}

As observed previously, the vacuum state $\ket{0}$ is a superposition of
states with even number of protons and neutrons. Therefore, since the
Hamiltonian is charge conserving, the SCQRPA eqs. \rf{3.1.10} split into
two equations, one for the charge-conserving excitations and the other for
the charge-exchange excitations.
Being interested only in investigating the
Fermi transitions, we will consider only the last ones,
\be
\Gamma^{\d}_{pn}=X_{pn}\frac{\A^{\d}_{pn}}{\sqrt{2\Omega}}
-Y_{pn}\frac{\A_{pn}}{\sqrt{2\Omega}},
\label{3.5.1}\ee
with
\be
\ket{pn}=\Gamma^{\d}_{pn}\ket{0}.
\label{3.5.2}\ee
The normalization condition
\be
\ov{pn}{pn}=\bra{0}[\Gamma_{pn}, \Gamma^{\d}_{pn}]\ket{0}=1,
\label{3.5.3}\ee
gives
\be
\left(X_{pn}^2-Y_{pn}^2\right)
\bra{0}[\A_{pn},\A^{\d}_{pn}]\ket{0}=2\Omega.
\label{3.5.4}\ee
Next, we introduce the normalized pair creation operators
\be
{\tilde \A}_{pn}^{\d}=D^{-1/2}_{pn} \frac{\A^{\d}_{pn}}{2\Omega},
\label{3.5.5}\ee
and the normalized amplitudes
\be
{\cal X}_{pn}=D^{1/2}_{pn}X_{pn},~~~~~~~~~ {\cal Y}_{pn}=D^{1/2}_{pn}Y_{pn},
\label{3.5.6}\ee
with
\be
D_{pn}= \bra{0}\frac{[\A_{pn},\A^{\d}_{pn}]}{2\Omega}\ket{0}
=1-\frac{\bra{0}\N_{pp}+\N_{nn}\ket{0}} {2\Omega},
\label{3.5.7}\ee
to write the SCQRPA equations as
\be
\mat{{\bar A}}{{\bar B}}{{\bar B}^*}{{\bar A}^*}\bin{{\cal X}_{pn}}{{\cal Y}_{pn}}=
\w\mat{1}{0}{0}{-1} \bin{{\cal X}_{pn}}{{\cal Y}_{pn}}.
\label{3.5.8}\ee
Here $\w$ is the SCQRPA excitation energy and
\br
{\bar A}&=&\bra{0}\left[{\tilde \A}_{pn},[{\cal H},{\tilde \A}_{pn}^{\d}]\right]\ket{0}
=\bra{0}\left[[{\tilde \A}_{pn},{\cal H}],{\tilde \A}_{pn}^{\d}]\right]\ket{0},
\nn\\
{\bar B}&=&-\bra{0}\left[{\tilde \A}_{pn},[{\cal H},{\tilde \A}_{pn}]\right]\ket{0}.
\label{3.5.9}\er
Using the expression for ${\cal H}$ in the quasiparticle basis, one can evaluate
the double commutators in eq. \rf{3.5.7}, and obtain
\br
{\bar A}&=&{\bar A}^{11}+{\bar A}^{22}+{\bar A}^{40},
\nn\\
{\bar B}&=&{\bar B}^{40}+{\bar B}^{22},
\label{3.5.10}\er
with
\br
{\bar A}^{ij}&=&\bra{0}\left[{\tilde \A}_{pn},[{\cal H}^{ij},{\tilde \A}_{pn}^{\d}]\right]
\ket{0}
\nn\\
{\bar B}^{ij}&=&-\bra{0}\left[{\tilde \A}_{pn},[{\cal H}^{ij},{\tilde \A}_{pn}]\right]\ket{0}.
\label{3.5.11}\er

The explicit expressions for ${\bar A}^{ij}$ and ${\bar B}^{ij}$ are:
\be
\bar{A}^{11}=E_p+E_n,
\label{3.5.12}\ee
with $E_a$ given by eq. \rf{3.2.5}, and
\br
\bar{A}^{22}&=&
\frac{1}{2}(F_{pn}\beta-G_{pn}\alpha)
+(F_{pn}\alpha-G_{pn}\beta)
\frac{\bra{0}\left(2\Omega-\N\right)^2\ket{0}}
{2\bra{0}\left(2\Omega-\N\right)\ket{0}}
\nn\\
&-&[4G_pu_p^2v_p^2+4G_nu_n^2v_n^2+F_{pn}\alpha-G_{pn}\beta]
\frac{\bra{0} \A_{pn}^{\d}\A_{pn}\ket{0}}
{\bra{0}\left(2\Omega-\N\right)\ket{0}}
\label{3.5.13}\\
&+&[G_p(u_p^4+v_p^4)+G_n(u_n^4+v_n^4)+F_{pn}\beta-G_{pn}\alpha]
\frac{\bra{0} \A_{pp}^{\d}\A_{pp}
-2\N_{pn}\N_{np}\ket{0}}
{\bra{0}\left(2\Omega-\N\right)\ket{0}}
\nn\\
&-&4(G_pu_p^2v_p^2+G_nu_n^2v_n^2) \frac{\bra{0}(2\Omega-\N)\N \ket{0}}
{\bra{0}\left(2\Omega-\N\right)\ket{0}},
\nn\er
\br
\bar{A}^{40}
&=&-(G_{pn}+F_{pn})u_pv_pu_nv_n
\frac{\bra{0}\A_{pn}^{\d}\A_{pn}^{\d}+\A_{pn}\A_{pn}\ket{0}}
{\bra{0}\left(2\Omega-\N\right)\ket{0}},
\label{3.5.14}\er
\be
\bar{B}^{40} =-(G_{pn}+F_{pn})u_pv_pu_nv_n
\left(1-\frac {\bra{0}\left(2\Omega-\N\right)^2
-2\A_{pn}^{\d}\A_{pn}\ket{0}}
{\bra{0}\left(2\Omega-\N\right)\ket{0}} \right),
\label{3.5.15}\ee
\br
\bar{B}^{22}&=&
-[4G_pu_p^2v_p^2+4G_nu_n^2v_n^2+F_{pn}\alpha-G_{pn}\beta]
\frac{\bra{0} \A_{pn}\A_{pn}\ket{0}}
{\bra{0}\left(2\Omega-\N\right)\ket{0}}
\nn\\
&+&[G_p(u_p^4+v_p^4)+G_n(u_n^4+v_n^4)+F_{pn}\beta-G_{pn}\alpha]
\frac{\bra{0} \A_{pp}\A_{nn} \ket{0}}
{\bra{0}\left(2\Omega-\N\right)\ket{0}},
\label{3.5.16}\er
where the short notations,
\be
\N=\N_{pp}+\N_{nn},
\label{3.5.17}\ee
and
\be
\alpha=u_p^2v_n^2+u_n^2v_p^2;~~~~~~~~~ \beta=u_p^2u_n^2+v_n^2v_p^2,
\label{3.5.18}\ee
have been used.

It easily can be seen that,
when the ground state correlations
in the SCQRPA equations are neglected, \ie when the SCQRPA vacuum $\ket{0}$
is replaced by the quasiparticle vacuum $\ket{BCS}$,
the usual QRPA equations
\br
A&=&\frac{\Omega}{2}(G_p +G_n)
-\beta G(pnpn;0) +\alpha F(pnpn;0),
\nn\\
B&=& [F(pnpn;0)- G(pnpn;0)]u_nv_nu_pv_p,
\label{3.5.19}\er
are recovered.

\renewcommand{\theequation}{\arabic{section}.\arabic{equation}}
\setcounter{equation}{0}

\section {Renormalized QRPA equations (RQRPA)}
\label{sec4}

As pointed out before, to solve the SCQRPA (SCRPA) equations exactly is very
demanding from a computational point of view, and this has not been done so
far in realistic cases.
Several authors
\cite{Toi95,Krm96,Sch96,Krm97,Toi97,Sim97,Mut97,Kar93,Ngu97}
have suggested different  approximations to derive simpler
equations to be amenable for the numerical calculations. In essence,
the simplifying hypothesis consists in factorizing the two-body correlations
in the SCQRPA (SCRPA) equations of motion. For a normal system this implies
to approximate the ground state two-body density by products of
one-body densities,
\be
\bra{0}c_a^{\d}c_b^{\d}c_d c_c\ket{0}
\cong \bra{0}c_a^{\d}c_c\ket{0} \bra{0}c_b^{\d}c_d\ket{0}
-\bra{0}c_a^{\d}c_d\ket{0} \bra{0}c_b^{\d}c_c\ket{0}.
\label{4.1}\ee
The parallel approximation in a superfluid system is:
\br
\bra{0}a_a^{\d}a_b^{\d}a_d a_c\ket{0}
&\cong& \bra{0}a_a^{\d}a_c\ket{0} \bra{0}a_b^{\d}a_d\ket{0}
-\bra{0}a_a^{\d}a_d\ket{0} \bra{0}a_b^{\d}a_c\ket{0}
\nn\\
&+&\bra{0}a_a^{\d}a_b^{\d}\ket{0} \bra{0}a_d a_c\ket{0},
\nn\\
\bra{0}a_a^{\d}a_b^{\d}a_d^{\d} a_c\ket{0}
&\cong& \bra{0}a_a^{\d}a_c\ket{0} \bra{0}a_b^{\d}a_d^{\d}\ket{0}
-\bra{0}a_a^{\d}a_d^{\d}\ket{0} \bra{0}a_b^{\d}a_c\ket{0}
\nn\\
&+&\bra{0}a_a^{\d}a_b^{\d}\ket{0} \bra{0}a_d^{\d} a_c\ket{0},
\nn\\
\bra{0}a_a^{\d}a_b^{\d}a_d^{\d} a_c^{\d}\ket{0}
&\cong& \bra{0}a_a^{\d}a_c^{\d}\ket{0} \bra{0}a_b^{\d}a_d^{\d}\ket{0}
-\bra{0}a_a^{\d}a_d^{\d}\ket{0} \bra{0}a_b^{\d}a_c^{\d}\ket{0}
\nn\\
&+&\bra{0}a_a^{\d}a_b^{\d}\ket{0} \bra{0}a_d^{\d} a_c^{\d}\ket{0},
\label{4.2}\er
and the corresponding hermitian conjugates.

The SCQRPA ground state does not have a fixed number of quasiparticles. Thus
its quasiparticle densities have both terms that conserve the number of
quasiparticles and terms that violate it. In field theory language,
one would say that in eq. \rf{3.5.1} we have approximated
the four-point function by a product of two-point functions.
Note that even with this approximation, to calculate the one-body densities
we have to evaluate the ground state $\ket{0}$ explicitly, and this is very
difficult in practice. Therefore, the one-body density (two-point function)
is approximated so that it is not required to determine
the excitations and the ground state simultaneously. This class of
approximations are known as RQRPA (RRPA).
They are valid as long as the effect of the two-body
correlations can be neglected, and their main advantage,
in comparison with SCQRPA (SCRPA), is
the numerical simplicity.

In this paper we derive the RQRPA equations for the O(5) model following the
scheme described above. The RQRPA limits for the expectation values are
shown in the Appendix B. Within this limit, the generalized BCS equations
\rf{3.4.5} reduce to
\br
&&\e_p^{(\sss SM)}-\lambda_p
-\frac{G_p\Omega}{2}\left[1 -\frac{\bra{0}\N_{pp}\ket{0}}{\Omega}\right]
(u_p^2-v_p^2)
\nn\\
&-&G_p\left[v_p^2+(u_p^2-v_p^2)
\frac{\bra{0}\N_{pp}\ket{0}}{2\Omega}\right]
\nn\\
&-&\frac{1}{2}(F_{pn}+G_{pn})\left[v_n^2+(u_n^2-v_n^2)
\frac{\bra{0}\N_{nn}\ket{0}}{2\Omega}\right]=0,
\label{4.3}\er
and an analogous equation for neutrons.

In the BCS limit the self-energy terms for protons are given by \rf{3.2.7},
\ie by $F(pppp;0)=-G_p$
times ${\rm N}_p/2\Omega$, plus $F(ppnn;0)=-(F_{pn}+G_{pn})/2$ times
${\rm N}_n/2\Omega$.
In the RQRPA one has to consider the average number of neutrons and protons
in the SCQRPA ground state $\ket{0}$ (see eq. \rf{3.4.2}). Thus, the
self-energy terms in the RQRPA have the same interpretation as in the BCS case.

Using the number condition \rf{3.4.2}, the eq. \rf{4.3} can be rewritten as
\be
\e_p^{(\sss SM)}-\lambda_p-G_p\frac{Z}{2\Omega} -(F_{pn}+G_{pn}) \frac{N}{4\Omega}
-\frac{{G}_p\Omega D_{pp}}{2} (u_p^2-v_p^2)=0,
\label{4.4}\ee
with
\be
D_{aa}=1-\frac{\bra{0}\N_{aa}\ket{0}}{\Omega}.
\label{4.5}\ee
Thus in the RQRPA limit the generalized BCS equations are equal to the
usual ones, except for the renormalization of the gap parameter by the
factor $D_{pp}$.

On the other hand, in this limit $\bar A$ gets contributions only from the
11 and 22 terms, \ie
$\bar A=\bar{A}^{11}+ \bar{A}^{22}$,
and  $\bar B$ only from the 40 term,\ie
$\bar B=\bar{B}^{40}$. One obtains
\br
\bar A&=&\frac{\Omega}{2}(G_p D_{pp} +G_n D_{nn})
-\beta G(pnpn;0)D_{pn} +\alpha F(pnpn;0)D_{pn},
\nn\\
\bar B&=& [F(pnpn;0)- G(pnpn;0)]D_{pn} u_nv_nu_pv_p,
\label{4.6}\er
where $\alpha$ and $\beta$ are given by \rf{3.5.18} and
$D_{pn}$ by \rf{3.5.7}.
Thus, we see that, compared with the QRPA equations \rf{3.5.16}, the matrix
elements
$F(pnpn;0)$ and $G(pnpn;0)$ are now renormalized by $D_{pn}$, as already
was known previously \cite{Toi95}.
Yet, our careful renormalization procedure shows that
the BCS quasiparticle energy
\begin{equation}
\varepsilon_a^{BCS}= \frac{\Omega G_a}{2}
=-\frac{G(aaaa;0)}{4},
\label{4.7}\end{equation}
is renormalized as
\begin{equation}
\varepsilon_a^R=\varepsilon_a^{BCS}D_{aa}.
\label{4.8}\end{equation}
This renormalization is a cooperative effect coming both from $\bar{A}^{11}$
and $\bar{A}^{22}$. In particular the former is a consequence of the coupling
of the mean field to excitations, embodied in the generalized BCS equations
\rf{4.3}.
\footnote{
To generalize the result  \rf{4.8} to the many level case, where
\[
\varepsilon_a^{BCS}
=-\frac{1}{4u_av_a}(2j_a+1)^{-1/2}\sum_b(2j_b+1)^{1/2}u_bv_bG(aabb;0),
\]
there are at least two possibilities:
\[
\varepsilon_a^R
=-\frac{D_{aa}}{4u_av_a}(2j_a+1)^{-1/2}\sum_b(2j_b+1)^{1/2}u_bv_bG(aabb;0),
\]
and
\[
\varepsilon_a^R
=-\frac{1}{4u_av_a}(2j_a+1)^{-1/2}\sum_b(2j_b+1)^{1/2}u_bv_bG(aabb;0)D_{ab},
\]
The last one has been obtained in ref. \cite{Krm97} from the self-consistency
between the residual interaction and the mean field. See also ref.
\cite{Mut97}.}

\setcounter{equation}{0}
\renewcommand{\theequation}{\arabic{section}.\arabic{equation}}

\section {Numerical results and discussion}
\label{sec5}

An exhaustive numerical study of the O(5) model has been performed for
different values of $N$, $Z$ and $\Omega$, as well as for a large set of
parameters.
For the discussion we have selected the following three samples:
\brn
{\rm (A)}:~\Omega&=&~5,~~N=~8,~Z=~2,~G_p=G_n=0.4~{\rm MeV},
~F_{pn}=0.4~{\rm MeV},
\nn\\
{\rm (B)}:~\Omega&=&10,~N=14,~Z=~6,~G_p=G_n=0.3~{\rm MeV},
~F_{pn}=0.3~{\rm MeV},
\nn\\
{\rm (C)}:~\Omega&=&25,~N=30,~Z=20,~G_p=G_n=0.2~{\rm MeV},
~F_{pn}=0.2~{\rm MeV}.
\ern
The strength $G_{pn}$, or more precisely the ratio,
\be
s=\frac{G_{pn}}{G_p},
\label{5.1}\ee
is kept as free parameter. The values of the single-particle energies are
of no relevance in the context of the present work and
we have simple chosen $\e_p=\e_n\equiv 0$. We believe that the last case
is as "realistic" as it can be within a schematic one-level model. Namely,
we mimic the calculation of a nucleus with $A=50$, within a rather large
configuration space, with the
Kisslinger and Sorensen \cite{Kis63} estimate for the pairing coupling
constant ($G\cong 25/A$ MeV), and the
strength $F_{pn}$ of the order of magnitude of $G_{p}$
($F_{pn}/G_{p}=1$).

\subsection {Solutions of the RPA equation}
\label{sec5.1}
By solving the equation \rf{3.5.8} we get
the frequencies $\omega$ and the amplitudes ${\cal X}$ and ${\cal Y}$.
From the last ones we can evaluate the $\beta^{\pm}$ strengths
for the transitions from the
even-even nucleus $(N,Z)$ to the odd-odd nuclei $(N\pm 1,Z\mp 1)$, \ie
\be
S^{\pm}=\sum_\l |\bra{0_\l}T_{\pm}\ket{0}|^2;
\label{5.2}\ee
where $T_+=N_{np}$ and  $T_-=N_{pn}$.
In an exact calculation the summation
in eq. \rf{5.2} goes over all states in the odd-odd nuclei.
But, within a QRPA-like calculation done here there is only one intermediate
state $\ket{0_{\sss int}}$, and
\br
\bra{0_{\sss int}}T_-\ket{0}&=& \sqrt{2\Omega{D}_{pn}}
({\cal X} u_pv_n+ {\cal Y} v_pu_n),
\nn\\
\bra{0_{\sss int}}T_+\ket{0}&=& \sqrt{2\Omega{D}_{pn}}
({\cal Y} u_pv_n+ {\cal X} v_pu_n).
\label{5.3}\er

It has been pointed out in ref. \cite{Hir96} that the sum rule
\be
S^{-}- S^{+}=N-Z,
\label{5.4}\ee
is violated in the RQRPA. However, it was shown
\cite{Krm96,Krm97,Del97} that this does not happens when both the correct
particle number condition
\rf{3.4.2} is used and only $J=0$ ground state pn
correlations are included. Under these circumstances
$\bra{0}\N_p\ket{0}$ equals
$\bra{0}\N_n\ket{0}$ and the above sum rule is fulfilled in RQRPA as well as
in SCQRPA.

In figures 1 and 2 are plotted the numerical results for $\omega$ and $S^-$,
respectively. We do not show the results for $S^+$ as they follow from
$S^-$ and the sum rule \rf{5.4}.
As expected, for $\omega$ the RQRPA results always fall in between those
of the QRPA and SCQRPA.
The QRPA collapses at a value of $s=s_{\rm crit} $ that decreases
when $\Omega$ is increased
($s_{\rm crit}= 2.10,~1.41$ and $1.10$, for the cases (A), (B) and (C),
respectively).
On the contrary, in RQRPA and SCQRPA there occurs no collapse and $\omega$
approaches zero asymptotically
when $s\go \infty$. From the figures we see that for
$s<s_{crit}$, all three approximations agree, but near the transition
point and beyond they differ. This behaviour is more evident in case (A).
The exact results for the transition strengths $S^-$ are also shown in
figure 2.
As evidenced from this figure, for $s\lsim s_{\rm crit}$ the approximate
$S^{\pm}$ strengths coincide so well with the exact ones, as well as among
themselves, that is hard to distinguish them visually.
When $s$ approaches $s_{\rm crit}$ the QRPA values of $S^{\pm}$ increase
very fast and go to infinity.
The SCQRPA results also reproduce nicely the exact results in the
neighborhood of the critical value of $s$ in all three cases. Moreover, in the
case A, SCQRPA yields approximately correct  $S^{\pm}$ strengths even for
values of $s$ very far from the
physical region for this parameter ($s\cong 1$).The variations of $z$, $\ex{\N}$ and $v_n^2$ in the case C, as
a function of $s$, are illustrated in figure 3.
We see that: i) the approximate calculations of the number of
quasiparticles in the ground state of the even-even nuclei only differ
from one another for $s \geq s_{\rm crit}$, and ii)
they undergo a sudden change for $s \simeq s_{\rm crit}$,
where the QRPA value goes to infinity. The variation of $v_{n}^{2}$
with $s$ in SCRPA and RQRPA is a direct consequence of the particle
number condition \rf {3.4.2} and is more pronounced in the
latter case.

\subsection {Excitation energies of the odd-odd nuclei}
\label{sec5.2}

While the RPA results for transition strengths $S^{\pm}$ can be directly
compared with the corresponding exact values, as well as with the
experimental data in a realistic case, this {\em does not} happens with
the pn-QRPA energies $\omega$. In fact,
one important question is to establish the relationship between
$\omega$ and the excitation energies $E^{\pm}$ of the intermediate
odd-odd nuclei $(N\pm 1,Z\mp 1)$ from the even-even nucleus $(N,Z)$.
This question is particularly relevant for the $\beta \beta$ decay.
Actually, in most
of the related theoretical studies done in the framework of the QRPA,
this problem
is circumvented by the use of the experimental ground state energies
of the initial, intermediate and final nuclei, and only
the relative spacing of the excitation energies is coming from the
roots of the
QRPA equation. (See, for instance, ref. \cite{Toi97}.)

In ref. \cite{Hir90} (see also ref. \cite{Krm97}) it was proposed that
the relationship between $\omega$ and $E^{\pm}$ is given by
\be
E^{\pm}=\w\mp\lambda_p\pm\lambda_n,
\label{5.5}\ee
in full analogy with the simple BCS case \cite{Rin80}.

However, from a more fundamental point of view, in the framework of the
SCQRPA, it is more natural to define these excitation energies as
\cite{Duk96}
\be
E^{\pm}_{\rm sc}=\w\mp
\bra{0}[\Gamma_{np},[\lambda_pN_p+\lambda_nN_n,\Gamma^{\d}_{np}]]\ket{0},
\label{5.6}\ee
which leads to
\be
E^{\pm}_{\rm sc}=\w\mp ({\cal X}^2+{\cal Y}^2)\sum_a\lambda_a(u_a^2-v_a^2).
\label{5.7}\ee

To test the quality of these two prescriptions we should remember the
following properties of the energies $E^{\pm}$:\\
1) for $\e_p=\e_n$ and $F_{pn}=0$, $E^-$ should pass through zero at $s=1$,
and\\
2) when the energy shift $\Delta=\e_p-\e_n$
(which can simulate the Coulomb energy displacement) is introduced,
$E^{\pm} \go E^{\pm}\mp \Delta$.

Obviously, in the exact calculation both of these conditions are fulfilled.
Additionally, from the eq. \rf{3.4.5}
and the fact that $\w$ does not change by a global shift of
the single particle energies,
we see that while the relation \rf{5.5}, being linear in
$\lambda_p-\lambda_n$, obeys the condition 2), the prescription \rf{5.7} does not.
On the other hand, from the results shown in figure 4 for the case (B)
with $F_{pn}=0$, we also conclude that only the first prescription for
$E^-$ is consistent with the condition 1). Thus, from now on we will
not discuss anymore the energies $E^{\pm}$ based on the eq. \rf{5.7}.
Why this last definition of the excitation energies leads
in our context to unphysical
consequences  will be investigated in a future work.

In figure 5 are compared the approximated energies $E^-$ with the exact ones.
All three approximations exhibit quite similar behavior for
$s\lsim s_{\rm crit}$.
Within this range of values for $s$, they also agree reasonable well with
the exact results. Yet, for
$s>s_{\rm crit}$ the agreement between the exact and the RQRPA and SCQRPA
results is not anymore so good and worsens when $\Omega$ is increased.
In any event the SCQRPA is preferable to the RQRPA, both close to the
critical point and beyond.

In all previous studies of the O(5) model \cite{Hir96,Sam97,Del97} only
the excitations in the $(N-1,Z+1)$ systems have been studied. But, the
energies $E^+$ could be as interesting for examining the different
approximations as are the former ones. Thus, we show them in figure 6.
The first issue that attracts attention, when comparing with the results
presented in figure 5, is that the discrepancies among different QRPA theories
for $E^+$, in the cases (A) and (B), are much more significant than for $E^-$.
The dissimilarities
with the exact results are also more pronounced, and especially in the
former example. We do not find out any immediate explanation for this outcome.
Yet, one should not be seriously worried by the results obtained
in the case (A), as in this example there is  only one proton in the final
state.
On the other hand, in the case (C) all three QRPA approaches for $E^+$
are quite
similar before the collapse of the QRPA, and reproduce reasonable well the
exact results. It should also be pointed out that
only the SCQRPA is capable to account for the minimum of the exact $E^+$.

\subsection{Double beta decay matrix elements}

The matrix element for the $\beta\beta_{2\nu}$ decay can be cast in the form:
\be
{\cal M}_{2\nu}=2\sum_\l
\frac{\bra{\bar{0}}T_-\ket{0_\l}\bra{0_\l}T_-\ket{0}}
{E^-_\l+\bar{E}^+_\l},
\label{5.8}\ee
where $\ket{0}$ and $\ket{\bar{0}}$ stand, respectively, for the
ground states in the initial
$(N,Z)$ and final $(N-2,Z+2)$ nuclei, with energies $E_0$ and $\bar{E}_0$,
and
\be
E^-_\l=E_\l-E_0;~~~~~~~~~
\bar{E}^+_\l=E_\l-\bar{E}_0,
\label{5.9}\ee
are the corresponding excitation energies from the ground states to
the virtual states $\ket{0_\l}$.

As said before, the matrix elements $\bra{0}T_-\ket{0_\l}$ obey the sum rule
\rf{5.4} with
$S^{\pm}$ given by \rf{5.2}. On the other hand, the matrix elements
$\bra{0_\l}T_-\ket{\bar{0}}$
fulfill the relation:
\be
\bar{S}^{-}- \bar{S}^{+}=N-Z-4,
\label{5.10}\ee
where
\be
\bar{S}^{\pm}=\sum_\l |\bra{0_\l}T_{\pm}\ket{\bar{0}}|^2.
\label{5.11}\ee

Within the QRPA calculations performed here there is only one intermediate
state. Thus
\be
{\cal M}_{2\nu}=2
\frac{\bra{\bar{0}}T_-\ket{0_{\sss int}}\bra{0_{\sss int}}T_-\ket{0}}
{E^-+\bar{E}^+},
\label{5.12}\ee
where, in the approximate calculations, $\bra{0_{\sss int}}T_-\ket{0}$ and
$E^-$ are given, respectively,  by eqs. \rf{5.3} and \rf{5.5},
\be
\bar {E}^+ =\bar{\w}-\bar{\lambda}_p+\bar{\lambda}_n,
\label{5.13}\ee
and
\be
\bra{\bar{0}}T_-\ket{0_{\sss int}}=
\sqrt{2\Omega\bar{D}_{pn}}
(\bar{{\cal Y}}\bar{u}_p\bar{v}_n+\bar{{\cal X}}\bar{v}_p\bar{u}_n).
\label{5.14}\ee
Here the barred quantities correspond to the final nucleus.

Before proceeding, let us note that we have found numerically that, within
the exact calculation, only one intermediate state contributes significantly.
This state is the isobaric analog state (IAS)  of
the state $\ket{0}$ in the isospin symmetry limit, \ie when $s=1$ and
$F_{pn}=0$. Thus, the eq. \rf{5.12} can also be
used in the exact calculations
for all practical purposes.
Moreover, we can write
\be
|{\cal M}_{2\nu}|=2
\frac{\sqrt{S^-\bar{S}^+}}
{E^-+\bar{E}^+}.
\label{5.15}\ee
Therefore, in the O(5) model the matrix element for the $\beta\beta_{2\nu}$
decay only depends on $S^-$, $E^-$, $\bar{S}^+$ and $\bar{E}^+$.
The results for the first two quantities have been discussed already, and
the results for the
remaining two are shown in figures 7 and 8. It should be specified that
the QRPA equations for the final systems collapse at significantly smaller
values of $s$, than for the initial ones.
Namely, now $s_{\rm crit}=1.17,~1.13$ and $1.04$, for the cases (A), (B) and
(C), respectively.
Nevertheless, from the comparison of figures
2 and 7 it can easily be seen that $\bar{S}^+$ is quite similar to
 $S^+$. In the same way from figures 6 and 8 one realizes that
$\bar{E}^+$ and
$E^+$ are alike to each other, although the discrepancies between the
exact and the approximates results are somewhat smaller for the former.
Finally, in figure 9 are presented the results for
the moments ${\cal M}_{2\nu}$.
Note that the sign of ${\cal M}_{2\nu}$ is not physically relevant.
Only for esthetic reasons we will choose it to be positive before
becoming zero and negative afterwards. From the comparison with the
exact calculations one can infer that
the all three approximate calculations account for the general behavior
of ${\cal M}_{2\nu}$, in the sense that they all cross the axis $s$ close
to the point where the exact calculations do it.
They, however overestimate its magnitude, for $s\le s_{\rm crit}$.
From eq. \rf{5.15} and the inspection of figures 2, 5, 7 and 8 it is easy
to discover that the main reason for the discrepancies with the exact
results comes from the differences in ${\bar E}^+$.
Close to the critical value of $s$ and beyond the SCQRPA is definitively
"better" than RQRPA.
Yet, in the case (C) all three approximations gives quite
satisfactory results.

\section {Summarizing Discussion}
\label{sec6}

In this work we again took up the presently very debated issue of extending
the quasiparticle RPA in the description of the charge-exchange transitions,
to account more accurately for the physics around the transition
point, which seems to be placed precisely in the region where the
particle-particle force takes its physical value.
An adequate tool to go beyond the QRPA approach seems to be its
self-consistent extension, \ie the SCQRPA.
This approach properly includes ground state correlations and the Pauli
principle, and has recently shown interesting
results for the pairing problem \cite{Duk96} as well as for other
physical systems \cite{Duk90,Kru94}.
Because of the complexity of the problem it seems appropriate that this
method be first tested on exactly solvable model cases before large
scale realistic calculations are performed.

It is considered
\cite{Hir96,Sam97,Del97}
that the schematic O(5) model
is the appropriate testing ground for the double beta decay, and
that it mimics rather accurately what is happening in realistic
model calculations.
But, one should be cautioned in doing such an extrapolation.
First, the O(5) model is only able to describe the F transitions that
play a marginal role in the $\beta\beta$ decay, in comparison with
the GT transitions. Second, the F processes are related with the isospin
conservation that is a good symmetry in nuclear physics, while the GT
processes are associated with the Wigner SU(4) symmetry \cite{Nak82,Ber90},
which is badly broken by the mean
field and only partially restored by the residual interaction.
Third, in realistic calculations the QRPA collapses in the physical region
for the particle-particle strength only in the GT case
\cite{Hir90,Krm94}.

To study specifically the GT transitions within a schematic
model one has to resort to the more general SO(8) model \cite{Eng88,Eva81}
where the $T=0$ and $T=1$ pairings coexist. Recently \cite{Eng97} it was
shown that the RQRPA, when applied
to this model, fails in reproducing the beta and double beta
decay properties beyond the phase transition point (SU(4) limit).
This failure was attributed to both: i) the need of changing the BCS basis,
due to the transition to a $T=0$ pairing region, and ii) the violation of the
Ikeda sum rule. One might hope that the SCQRPA would solve the problem.
Yet, although SO(8) is only slightly more complicated than O(5), we do not
know how to build up explicitly the correlated ground state for the former,
and, as was shown in the present work, this is the key point in solving
the SCQRPA equation.
The determination of the
correlated ground state is also the main difficulty in applying the
SCQRPA to real nuclei. Therefore, from the practical point of view, one of
the most important tasks to solve the SCQRPA equations is the development
of physically sound and numerically accurate approximate methods,
which do not require the explicit construction of the correlated
ground state. Schematic models such as the O(5) and SO(8) are perfect
testing grounds for this trial.

In the present paper we have studied in detail the properties of the
QRPA, RQRPA and SCQRPA  equations in the O(5) model.
Compared with the previous work \cite {Del97}, we have made several
improvements, namely: i) we have solved exactly the SCQRPA equations of
the O(5) model by evaluating everything exactly with the QRPA vacuum
state \rf{3.3.2}
ii) we correctly took into account the coupling of the mean-field
to the excitations.
We also have examined carefully the RRPA limit of the model. The importance
of all these features is clearly born out by the results.
The energies, transition strengths, and double beta decay matrix elements
are accounted for reasonably well.
It clearly turns out that the SCQRPA approach performs better than either
standard QRPA or renormalized QRPA for the description of these observables.
Only in the case (C) the QRPA collapses in the physical region for the
parameter $s$ ($s\cong 1$). Thus, the samples (A) and (B) mimic better the F
double beta decays and the example (C) the GT ones.

Before the collapse of the QRPA all three approaches reproduce well the exact
results. Near the transition point only the SCQRPA values are close to
the exact ones. Beyond that point both the SCQRPA and RQRPA
yield different results from the exact ones, but the
former are always qualitatively superior. It can be suspected that
this situation even prevails to some extent in the realistic
situation. On the other hand, it is very well known \cite{Rin80} that after
the point
where standard RPA breaks down, one has to change the single particle basis.
In the present situation, where the proton-neutron correlations become
``soft'', this implies that after the phase transition point one has to
consider the quasi-particles as mixed proton-neutron states.

Elaborating a SCQRPA in this new basis should allow to go in a smooth
way through the phase transition (as happens in other models
\cite{Duk90}), and we expect that then the SCQRPA will reproduce the exact
results not only before but also beyond the transition point.
In the same line one then also should
include proton-proton and neutron-neutron pair excitation terms
in the QRPA operator using the generalized Bogoljubov technique.
Unfortunately such a generalization of the theory is a nontrivial task
even in the O(5) model and demands considerable formal developments.
We intend to elaborate on this in the future.

A novel feature that has emerged from this work is that in the RRPA limit
not only the proton-neutron components of the
interaction are renormalized, as was well known previously \cite{Toi95}, but
also the quasi-particle energies. To get this effect it is essential
to consider the coupling of the excitation to the mean-field, embodied
in the generalized BCS equations \cite{Duk90}.

One further technical
aspect that deserves mentioning here concerns the inclusion of
so-called anomalous terms $a^{\dagger}a$ in the excitation
operator, as this was proposed in \cite{Duk98,Duk96}. Inclusion of such terms is
very important to fulfill both the energy weighted and energy non weighted
sum rules, as this was shown in \cite{Duk98} for the case of homogeneous
infinite nuclear matter.
A straightforward calculation also shows that such terms are essential to
fulfill the sum rule for F and GT transitions in
realistic finite nuclei.
Inclusion of these terms is also relevant for the correct
treatment of symmetries and the appearance of the Golstone mode \cite{Duk98}.
But, contrary to the infinite matter case, the consistent inclusion
of such terms in group theoretical models, like the one treated here,
is a delicate problem and demands further studies.

A particular feature of our model relates to the fact that the QRPA ground
state wave function could be constructed explicitly and therefore the
system of equations could be closed without further approximations.
In the realistic situation it is certainly far from obvious to obtain
the ground state wave function and approximations must be used to close
the system. In this respect efficient methods have been developed
in the past, like e.g. the number operator method \cite{Row68}.
On the other hand, it has been shown that the system of equations
can be closed without knowing the wave function
in coupling the SCRPA to the one body Green's function \cite{Duk98}.
This way was considered in eq. (16) of ref. \cite{Duk96}.
In the future the elaboration of a more consistent single particle
equation coupled to SCRPA will be a key point in closing
the equations also for realistic situations.

This work was supported in part by the DGICYT (Spain)
under contract No. PB 95/0123-C02-01.

\pagebreak
\renewcommand{\theequation}{\thesection.\arabic{equation}}
\appendix
\setcounter{equation}{0}
\begin{center}
\subsection*{Appendix A: Expectation values within the SCQRPA}
\end{center}
\setcounter{section}{1}

Except for the operators $\A_{pp}^{\d}\A_{nn}^{\d}$,
 $\A_{pp}^{\d}\A_{pp}$ and $\N_{np}\N_{pn}$,
the calculation of the expectation values can be easily done using the
completeness properties of the SCQRPA exchange mode and the properties
of the basis states.
For these three operators
we had to use the generator state technique from ref. \cite{Das73}.

Using the notation,
\be
\ex{\N}\equiv\bra{0}\N_{pp}\ket{0}=
\bra{0}\N_{nn}\ket{0},
\label{A.1}\ee
and the ground state given by \rf{3.3.2} we get
\be
\ex{\N}=
2N_0\sum_{l=0}^{\Omega}l\alpha_l^2z^{2l}\ov{2l}{2l},
\label{A.2}\ee
\be
\bra{0}\N_{pp}^2\ket{0}= \bra{0}\N_{nn}^2\ket{0}=
\bra{0}\N_{nn}\N_{pp}\ket{0}=4N_0\sum_{l=0}^{\Omega}l^2\alpha_l^2z^{2l}\ov{2l}{2l},
\label{A.3}\ee
\be
\bra{0}\A_{pp}^{\d}\A_{nn}^{\d}\ket{0}=
-\frac{4N_0}{2\Omega-1}\sum_{l=1}^{\Omega}l(2\Omega-2l+1)\alpha_l^2z^{2l-1}
\ov{2l}{2l},
\label{A.4}\ee
\be
\bra{0}\A_{pp}^{\d}\A_{pp}\ket{0}= \bra{0}\A_{nn}^{\d}\A_{nn}\ket{0}
=\frac{4N_0}{2\Omega-1}\sum_{l=1}^{\Omega}l(2l-1)\alpha_l^2z^{2l} \ov{2l}{2l},
\label{A.5}\ee
\be
\bra{0}\A_{pn}^{\d}\A_{pn}\ket{0}=\frac{2z^2}{1-z^2}(\Omega-\ex{\N}),
\label{A.6}\ee
\be
\bra{0}\A_{pn}\A_{pn}\ket{0}=
\bra{0}\A_{pn}^{\d}\A_{pn}^{\d}\ket{0}=\frac{2z}{1-z^2}(\Omega-\ex{\N}),
\label{A.7}\ee
and
\be
\bra{0}\N_{np}\N_{pn}\ket{0}= \bra{0}\N_{pn}\N_{np}\ket{0}
=2\ex{\N} -\bra{0}\A_{pp}^{\d}\A_{pp}\ket{0}.
\label{A.8}\ee

\newpage

\setcounter{equation}{0}
\begin{center}
\subsection* {Appendix B: RQRPA limit for the expectation values}
\end{center}
\setcounter{section}{2}
With
$\ex{\N}$ defined in \rf{A.1} we obtain
\be
\bra{0}\N_{pp}^2\ket{0}=
\bra{0}\N_{nn}^2\ket{0}
=\ex{\N}^2+\ex{\N}\left(1-{\ex{\N}\over 2\Omega}\right),
\label{B.1}\ee
\be
\bra{0}\N_{nn}\N_{pp}\ket{0}= \ex{\N}^2,
\label{B.2}\ee
\be
\bra{0}\A_{pp}^{\d}\A_{nn}^{\d}\ket{0}=
\bra{0}\A_{pn}^{\d}\A_{pn}^{\d}\ket{0}=
\bra{0}\A_{pn}\A_{pn}\ket{0}=0,
\label{B.3}\ee
\be
\bra{0}\A_{pp}^{\d}\A_{pp}\ket{0}=
\bra{0}\A_{nn}^{\d}\A_{nn}\ket{0}=
2\bra{0}\A_{pn}^{\d}\A_{pn}\ket{0}=
\frac{\ex{\N}^2}{\Omega},
\label{B.4}\ee
and
\be
\bra{0}\N_{np}\N_{pn}\ket{0}=
\bra{0}\N_{pn}\N_{np}\ket{0}=\ex{\N}
-\frac{\ex{\N}^2}{2\Omega}
=\ex{\N}\left(1-{\ex{\N}\over 2\Omega}\right).
\label{B.5}\ee
The following relations are also useful:
\be
\bra{0}\left(2\Omega-\N\right)^2\ket{0}
=4\Omega^2+ \frac{\ex{\N}}{\Omega}(4\Omega-1)(\ex{\N}-2\Omega),
\label{B.6}\ee
\be
\bra{0}\left(\Omega-\N_{pp}\right)^2\ket{0}
=\Omega^2+ \frac{\ex{\N}}{2\Omega}(2\Omega-1)(\ex{\N}-2\Omega),
\label{B.7}\ee
\br
\bra{0}2\left(\Omega-\N_{pp}\right)^2
-\A_{pp}^{\d}\A_{pp}\ket{0}&=&
2\left(\Omega^2+ \ex{\N}(1 -2\Omega)+ \frac{\ex{\N}^2}{\Omega}
(\Omega-1)\right).
\label{B.8}\er

\begin{figure}[t]
\begin{center} { \large Figure Captions} \end{center}
\caption{Frequencies $\w$ for the $(N,Z)$ systems for the cases (A), (B) and
(C), as a function of the particle-particle coupling constant $s$.}
\label{figure1}

\bigskip

\caption{Transition strengths  $S^-$ for the $(N,Z)$ systems for the cases
(A), (B) and (C), as a function of the particle-particle coupling constant $s$.}
\label{figure2}

\bigskip

\caption{The ratio $z={\cal Y}/{\cal X}$ (upper panel), expectation value of
the number of quasiparticles in the ground state $\ex{\N}=\bra{0}\N_p\ket{0}
=\bra{0}\N_n\ket{0}$ (middle  panel), and the occupation probability for
neutrons $v_n^2$ (lower panel). Within the QRPA, $\ex{\N}=2{\cal Y}^2$.}
\label{figure3}

\bigskip

\caption{Excitation energies $E^-$ for the case (B),
as a function of the particle-particle coupling constant $s$.
The approximate results were obtained with the
prescriptions \protect \rf{5.5} (upper panel) and \protect \rf{5.7}
(lower panel).}
\label{figure4}

\bigskip

\caption{Excitation energies $E^-$ for the cases (A), (B) and (C),
as a function of the particle-particle coupling constant $s$.
The approximate results were obtained from
eq. \protect \rf{5.5}.}
\label{figure5}

\bigskip

\caption{Excitation energies $E^+$ for the cases (A), (B) and (C),
as a function of the particle-particle coupling constant $s$.
The approximate results were obtained from
eq. \protect \rf{5.5}.}
\label{figure6}

\bigskip

\caption{Transition strengths  $\bar{S}^+$ for the $(N-2,Z+2)$ systems for
the cases (A), (B) and (C), as a function of the particle-particle coupling
constant $s$.}
\label{figure7}

\bigskip

\caption{Excitation energies ${\bar E}^+$ for the cases (A), (B) and (C),
as a function of the particle-particle coupling constant $s$.
The approximate results were obtained from
eq. \protect \rf{5.13}.}
\label{figure8}

\bigskip

\caption{Matrix element ${\cal M}_{2\nu}$
as a function of the particle-particle coupling constant $s$.}
\label{figure9}
\end{figure}
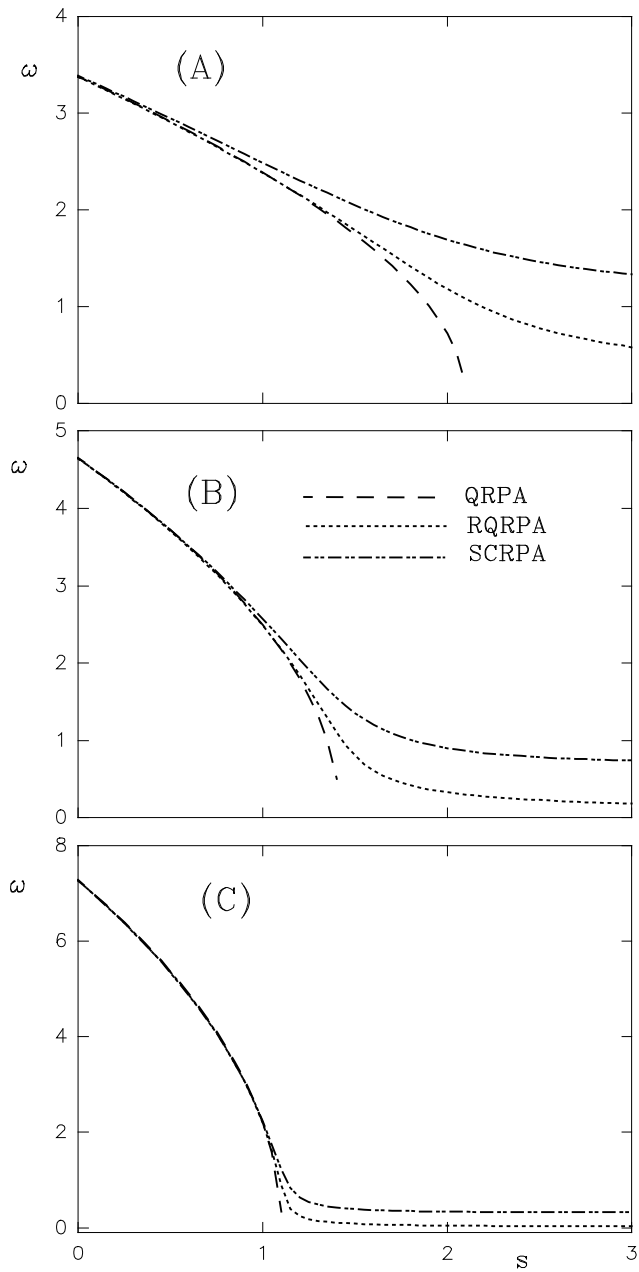
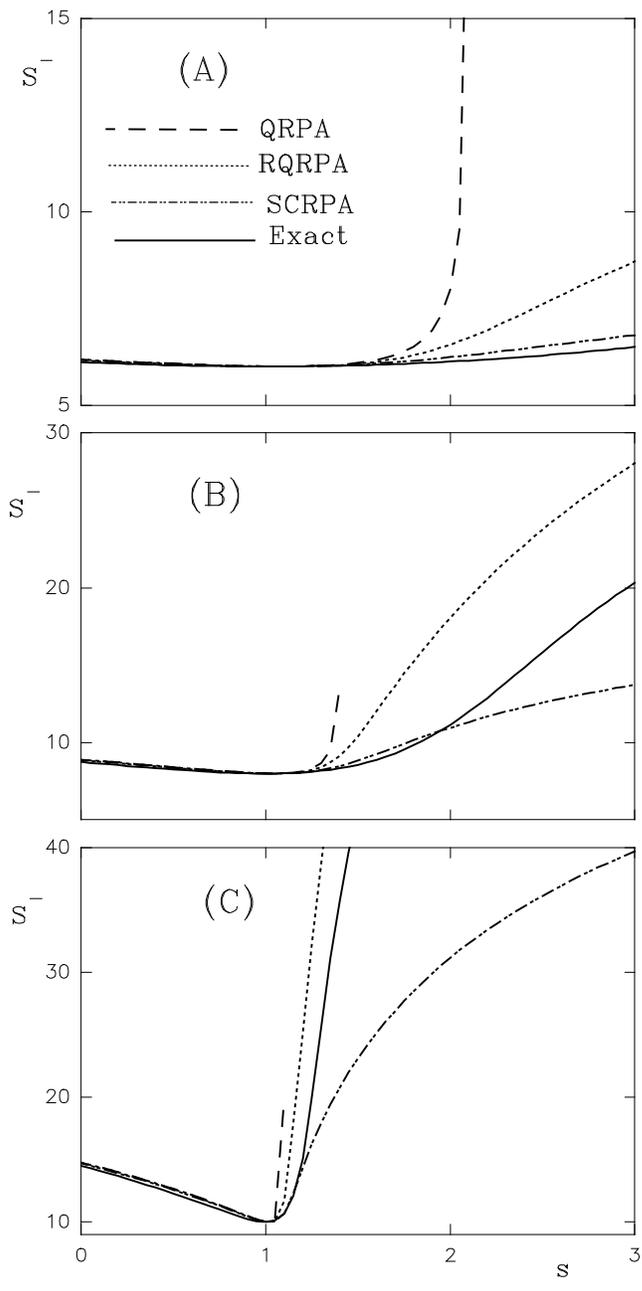
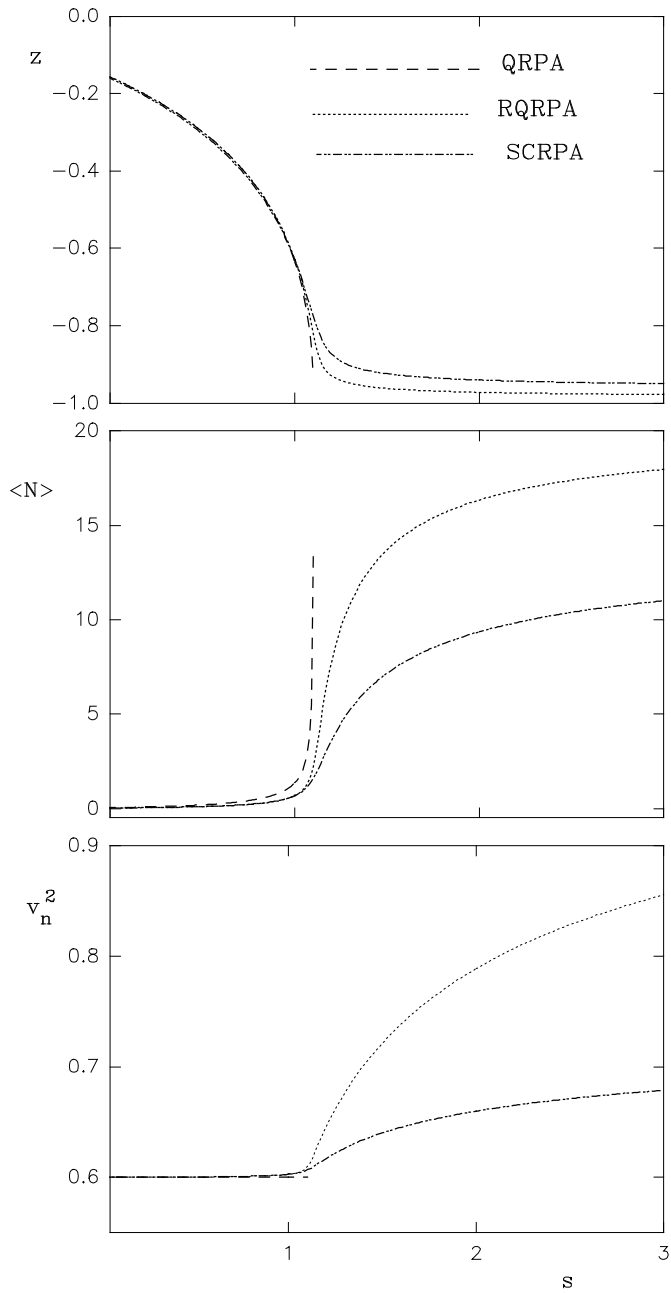
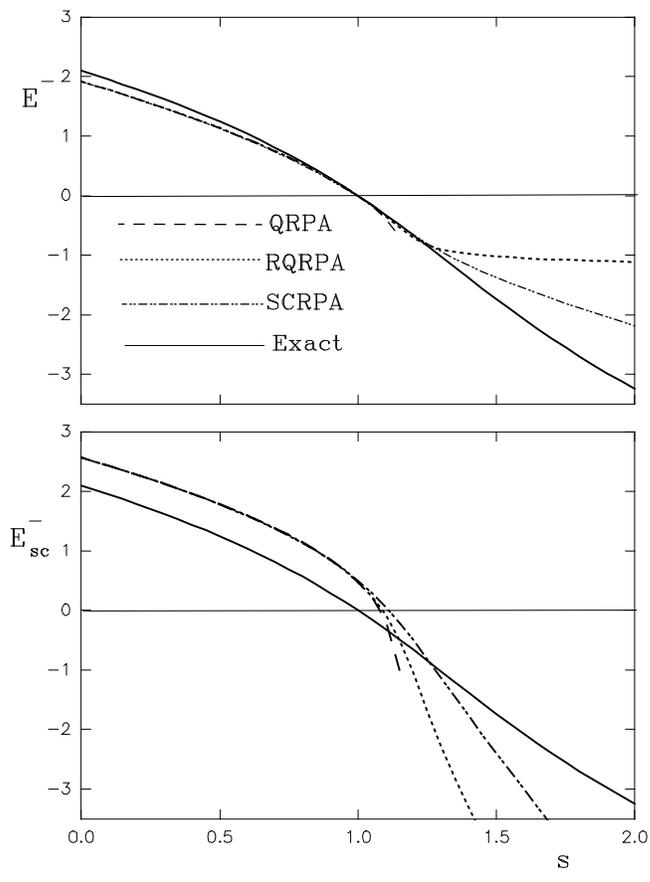
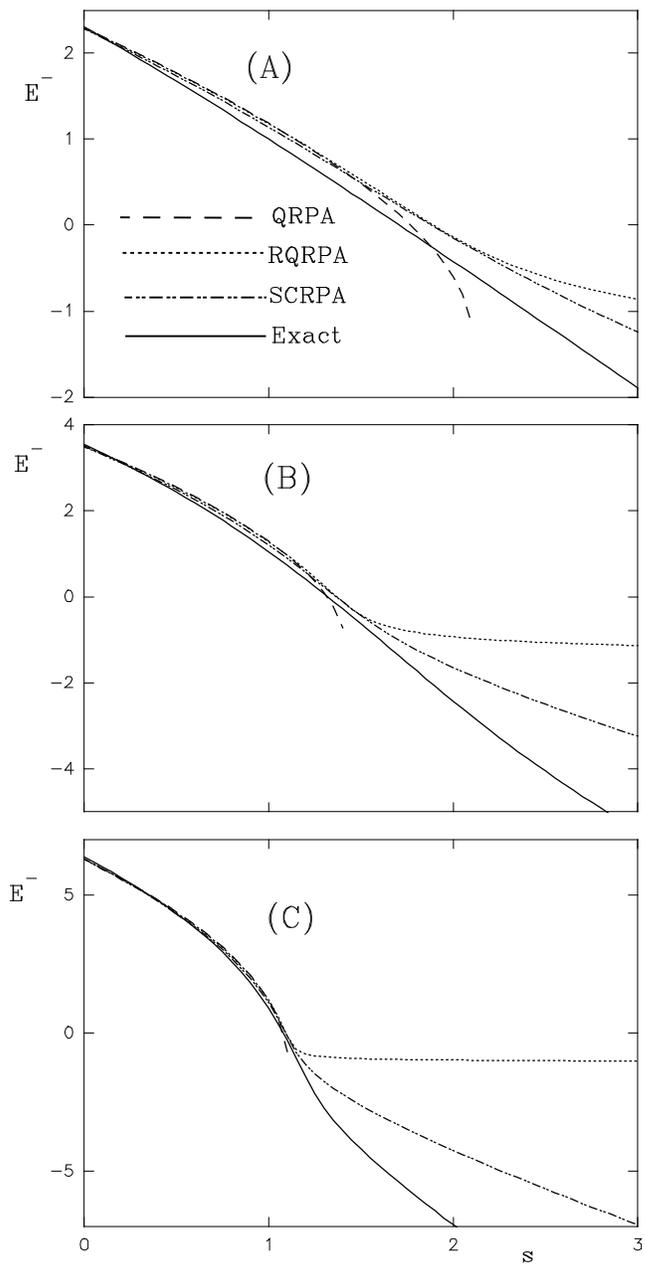
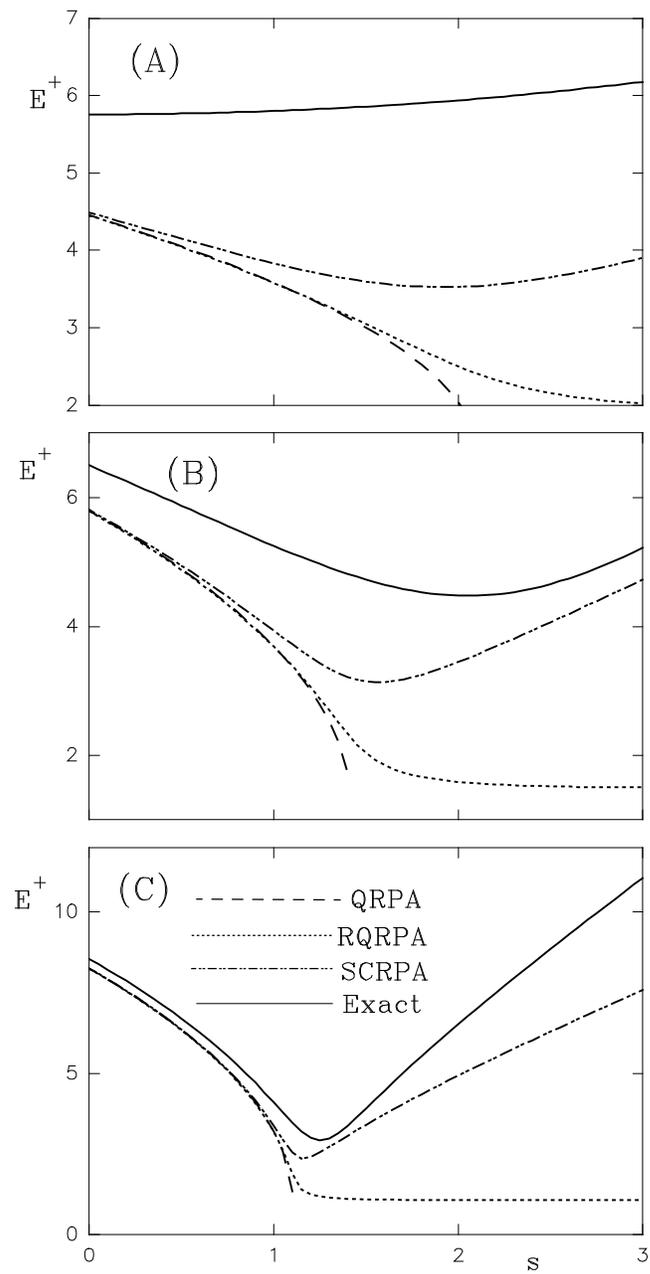
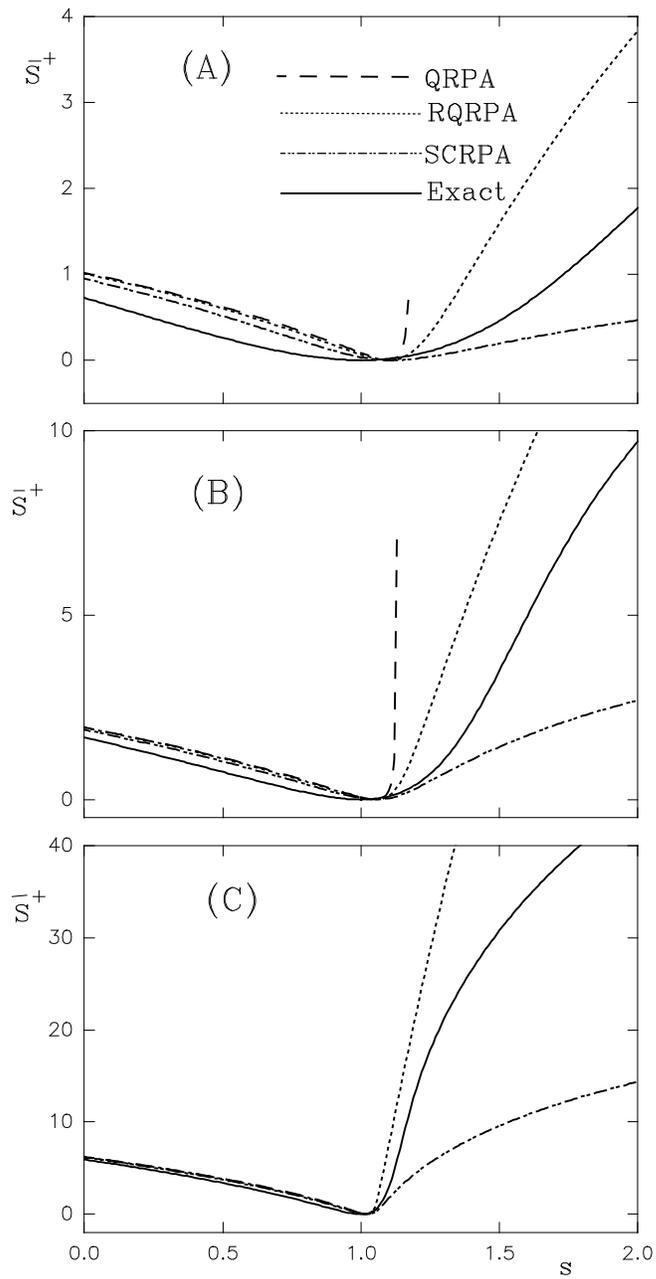
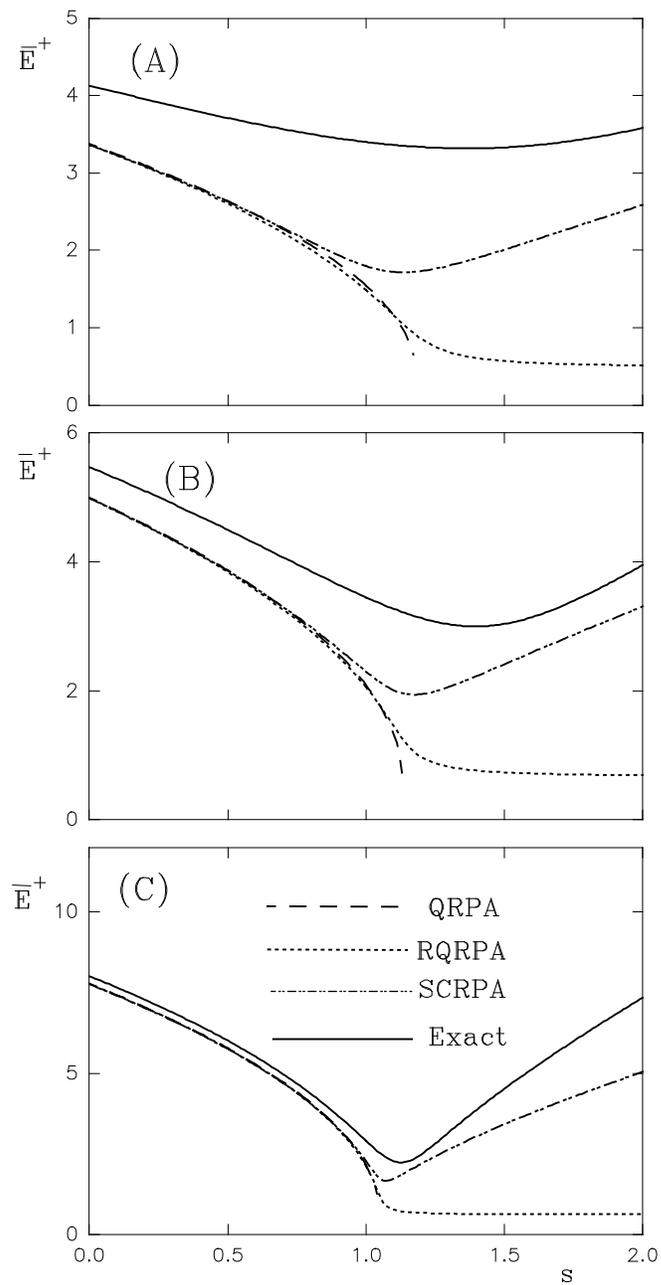
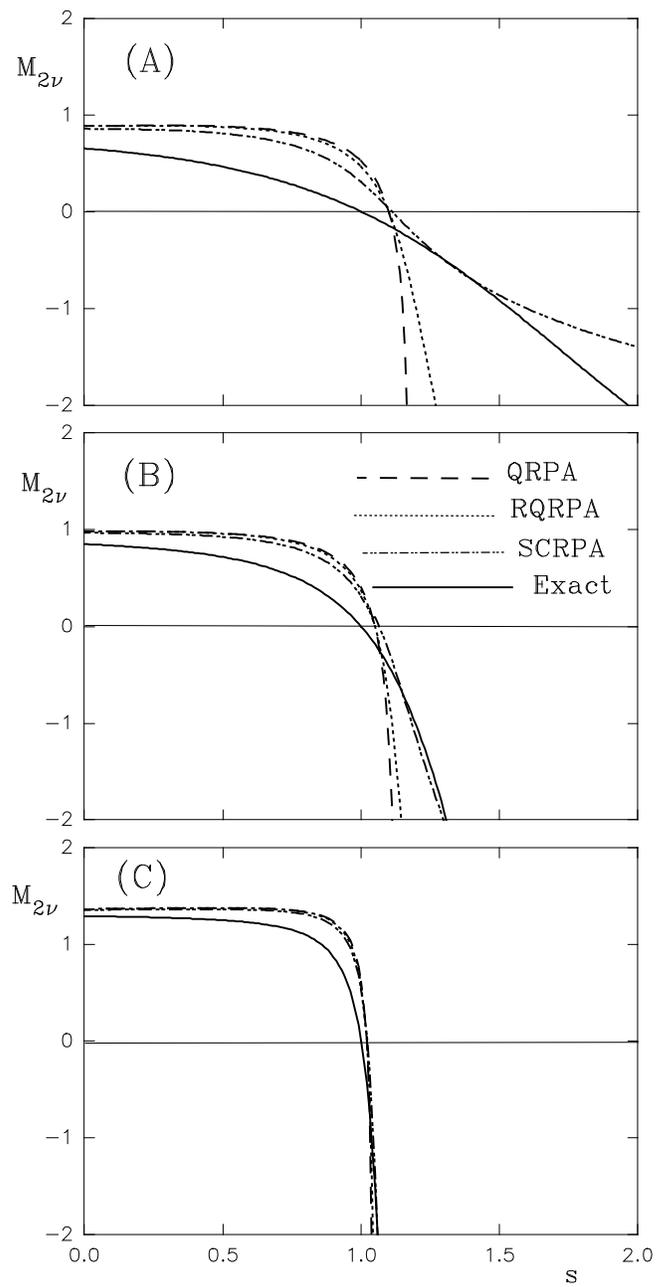
\end{document}